\documentclass{article}

\usepackage{PRIMEarxiv}

\usepackage[utf8]{inputenc} 
\usepackage[T1]{fontenc}    
\usepackage{hyperref}       
\usepackage{url}            
\usepackage{booktabs}       
\usepackage{amsfonts}   
\usepackage{amsmath}      
\usepackage{braket} 
\usepackage{nicefrac}       
\usepackage{microtype}      
\usepackage{lipsum}
\usepackage{fancyhdr}       
\usepackage{graphicx}       
\graphicspath{{media/}}     

\title{Shortcut to Adiabatic Isomeric Population Transfer of the $^{229}\mathrm{Th}$ Nucleus via Hyperfine Electronic Bridge}

\author{
Bo Liu\\
School of Information and Electrical Engineering, Hangzhou City University, Hangzhou 310015, China \\
Wu Wang
\thanks{Corresponding author: E-mail:~wangwu531@hainanu.edu.cn}\\
Center for Theoretical Physics \& School of Physics and Optoelectronic Engineering, \\
Hainan University, Haikou 570228, China\\
Institute of Theoretical Physics, Chinese Academy of Sciences, 
Beijing 100190, China\\
Yong Li
\thanks{Corresponding author: E-mail:~yongli@hainanu.edu.cn}\\
Center for Theoretical Physics \& School of Physics and Optoelectronic Engineering, \\
Hainan University, Haikou 570228, China\\
}

\begin{document}
\maketitle

\begin{abstract}
The $^{229}$Th nucleus is well known for its exceptionally low-lying nuclear isomeric level, which provides a unique platform for exploring electron-nucleus interactions and gives rise to a variety of rich physical phenomena. One such phenomenon is the hyperfine electronic bridge, which has recently been shown to enable efficient and precise manipulation of the nuclear isomeric levels of $^{229}$Th [W. Wang $et~ al.$, Phys. Rev. Lett. \textbf{133}, 223001 (2024)]. However, that study used the stimulated Raman adiabatic passage method, which requires relatively long operation times. In this work, we employ the stimulated Raman shortcut-to-adiabatic passage method, which dramatically shortens the operation time from the order of hundreds of milliseconds to hundreds of microseconds while maintaining a transfer efficiency of about $79.38\%$.
\end{abstract}


\section{Introduction}
The $^{229}$Th nucleus has an isomeric state with an extremely low energy of about 8 eV~\cite{PhysRevLett.132.182501,PhysRevLett.133.013201,Zhang2024}, enabling the development of a nuclear clock controlled by state-of-the-art vacuum ultraviolet lasers. Such a clock has been proposed for the detection of variations in the fundamental constants and dark matter~\cite{Peik2003,PhysRevLett.108.120802,Thirolf2024,Fuchs2025,Peik2021}. Moreover, the $^{229}$Th nucleus provides a unique platform to investigate fundamental electron-nucleus interactions, including the recollision induced nuclear excitation~\cite{PhysRevLett.127.052501,PhysRevC.106.024606,PhysRevResearch.5.043232}, the nuclear hyperfine mixing~\cite{Lyuboshitz_1966,Szerypo1990,Wycech1993,PhysRevC.57.3085,PhysRevC.64.064301,PhysRevLett.112.062503,PhysRevC.94.014323,PhysRevLett.128.043001,PhysRevResearch.5.023134,Wang2024,PhysRevLett.133.152503}, nuclear excitation by electron capture~\cite{PhysRevA.73.012715,PhysRevLett.127.042501,PhysRevLett.130.112501,Zhao2024}, and the electronic bridge (EB)~\cite{Batkin1979,Tkalya1992,PhysRevLett.105.182501,MULLER201784,PhysRevC.100.044306,PhysRevLett.125.032501,PhysRevC.102.024604,Li2023,PhysRevLett.124.192502}.

The EB process refers to a nuclear transition induced by a laser-driven electronic transition, in which the coupling between the nucleus and the laser field is effectively enhanced \cite{Tkalya1992}. In particular, a two-photon EB scheme enables the excitation of $^{229}$Th using two laser fields of around 300 nm \cite{Campbell2009}, thereby eliminating the need for a laser at about 150 nm. However, most EB studies have not accounted for hyperfine structure \cite{Batkin1979,Tkalya1992,PhysRevLett.105.182501,MULLER201784,PhysRevC.100.044306,PhysRevLett.125.032501,PhysRevC.102.024604,Li2023}. As the uncertainty of the isomeric energy has already been reduced to the kilohertz level, such neglect has become unreasonable. Additionally, the treatment of EB is in a perturbative way, and the decay of the excited states involved in two-photon EB is typically ignored.  Recently, these issues were addressed by introducing the hyperfine EB process and using a quantum-optical method in Ref.~\cite{Wang2024-2}. In that work,  efficient preparation of the $^{229}$Th isomer was achieved by using the stimulated Raman adiabatic passage (STIRAP) method. Nevertheless, the use of STIRAP technique rendered the preparation relatively slow.

In this work, we employ the stimulated Raman shortcut-to-adiabatic passage (STIRSAP)~\cite{PhysRevLett.109.100403,Du2016} in the quantum six-level system of $^{229}$Th$^{3+}$ ions introduced in Ref.~\cite{Wang2024-2}. Numerical simulations demonstrate that the STIRSAP method achieves efficient population transfer to the isomeric state within hundreds of microseconds. This duration is only a few hundredths of that required by the STIRAP scheme in Ref.~\cite{Wang2024-2}. Under identical operating conditions and pulse peak intensities, the STIRSAP method exhibits significantly higher transfer efficiency than STIRAP. In addition, the influence of one-photon detuning on the isomeric population transfer efficiency of STIRSAP is investigated. Lastly, compared with STIRAP, the STIRSAP method is  more robust against uncertainties in the two-photon detuning.

\section{Method}
\subsection{Hyperfine EB in $^{229}\mathrm{Th}$}
The combined nucleus-electron system is coupled through the hyperfine interaction. This system has a series of hyperfine levels characterized by the quantum numbers $(I, \Gamma J, F)$, where $I$ denotes the nuclear spin, $J$ is the electronic angular momentum, $F$ is the total angular momentum coupled by $I$ and $J$, and $\Gamma$ represents all other electronic quantum numbers. Each hyperfine level $(I, \Gamma J, F) $ is described by a dressed hyperfine state $\ket{[I \Gamma J] Fm_F}$, which is an entangled state involving both nuclear and electronic degrees of freedom \cite{Wang2024-2}. By using perturbation theory, $\ket{[I \Gamma J] Fm_F}$ can be expanded by means of hyperfine coupled basis $\ket{I \Gamma J, F m_F}$
\begin{equation}
| [I \Gamma J] F m_F \rangle =  |I \Gamma J, F m_F\rangle + \sum_{\tau K n } (-1)^{I+J_n+F} 
\begin{Bmatrix} I_n & J_n & F \\ J & I & K \end{Bmatrix} \frac{\langle \Gamma_n J_n \| 
T^{(\tau K)} \| \Gamma J \rangle}{E_{(I, \Gamma J)} - E_{(I_n, \Gamma_n J_n)}} \langle I_n 
\| M^{(\tau K)} \| I \rangle|I_n\Gamma_n J_n, F m_F\rangle,
\label{eq:perturbation}
\end{equation}
where $|I \Gamma J, F m_F\rangle$ denotes the leading hyperfine-coupled basis state with energy $E_{(I, \Gamma J)}$, and $|I_n\Gamma_n J_n, F m_F\rangle$ represents different hyperfine-coupled basis state with energy $E_{(I_n, \Gamma_n J_n)}$. The specific expression of hyperfine-coupled basis $|I \Gamma J, F m_F\rangle$ can be found in Ref.~\cite{Wang2024-2}. Here, $M^{(\tau K)}$ denotes the nuclear transition operator and $T^{(\tau K)}$ represents the electronic operator associated with the hyperfine interaction \cite{Wang2025}. Both are spherical tensor operators of rank $K$, with $\tau$ distinguishing between electric ($\tau = E$) and magnetic ($\tau = M$) types.

For the $^{229}\mathrm{Th}$ nucleus, the summation over nuclear spin $I_n$ in the above equation can be restricted to the ground level with nuclear spin $I_g$ and the isomeric level with nuclear spin $I_e$ due to the large energy gaps between these two states and all other nuclear levels. The dominant transition between the nuclear ground and isomeric levels of  $^{229}\mathrm{Th}$ is magnetic-dipole ($M1$) transition. Therefore, in the calculation of dressed  hyperfine states, the reduced matrix $\langle I_e \| M^{(\tau K)} \| I_g \rangle$ is required, which is determined by the reduced nuclear transition probability $B(M1,I_e\rightarrow I_g)=3|\langle I_e \| M^{(M1)} \| I_g \rangle|^2/[4\pi(2I_e+1)]$. In this work, a value of $B(M1,I_e\rightarrow I_g)=0.022$~W.u. is adopted~\cite{PhysRevLett.132.182501}. The detailed numerical method can be found in the Supplemental Material of Ref.~\cite{Wang2024-2}.

The hyperfine EB process describes a transition between two different hyperfine levels, e.g., $(I_g, \Gamma_1 J_1, F_g)\rightarrow (I_e, \Gamma_2 J_2, F_e)$, in which both the nuclear and electronic structures are changed \cite{Wang2025_2}. Compared with the conventional description of EB process~\cite{Batkin1979,Tkalya1992,PhysRevLett.105.182501,MULLER201784,PhysRevC.100.044306,PhysRevLett.125.032501,PhysRevC.102.024604,Li2023,PhysRevLett.124.192502}, the physical picture of the hyperfine EB is simplified as a two-level transition, where the numerous intermediate levels are effectively eliminated. Another advantage of hyperfine EB is that it allows for a straightforward construction of quantum-optical models involving only a few levels, opening the possibility of employing different quantum-control techniques \cite{Wang2024-2}. In this work, we focus on $^{229}\mathrm{Th}^{3+}$ ions, which has been proposed as a candidate system for nuclear clock \cite{Peik2003}. We emphasize that hyperfine EB transitions generally exist in different $^{229}\mathrm{Th}$ ions (and other nuclear system). Consequently, our quantum-optical approach is flexible and can be applied to other $^{229}\mathrm{Th}$ ions, such as  $^{229}\mathrm{Th}^{6+}$ \cite{Yu2025}, $^{229}\mathrm{Th}^{32+}$, and $^{229}\mathrm{Th}^{40+}$. For $^{229}\mathrm{Th}^{32+}$ and $^{229}\mathrm{Th}^{40+}$, it has been demonstrated that hyperfine EB transitions can significantly enhance the sensitivity to variations of the fine-structure constant \cite{Wang2025_2}.

\begin{figure}[t!]
\centering
\includegraphics[width=6cm]{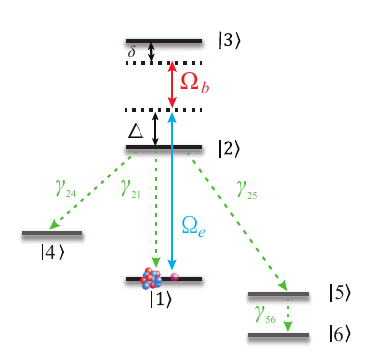}
\caption{\label{fig:2025_10_11_1} {The six-level model for $^{229}$Th$^{3+}$ ions begins with $|1\rangle \equiv |[I_g 7s_{1/2}]F=2\rangle$ and aims to  excite the isomeric state $|3\rangle \equiv  |[I_e 7s_{1/2}]F=2\rangle$ via the intermediate state $|2\rangle \equiv |[I_g 7p_{1/2}] F=3\rangle$. An electron-assisting laser (with Rabi frequency $\Omega_e$) drives the transition between $\ket{1}$ and $\ket{2}$, while a bridge-assisting laser (with Rabi frequency  $\Omega_b$) couples $\ket{2}$ and $\ket{3}$. The decay from $|2\rangle$ to $|1\rangle$ (with decay rate $\gamma_{21}$), $|4\rangle \equiv |[I_g 7s_{1/2}] F=3\rangle$ (with decay rate $\gamma_{24}$), and $|5\rangle \equiv |I_g, 6d_{3/2}\rangle$ (with decay rate $\gamma_{25}$), as well as the decay from $|5\rangle$ to $|6\rangle \equiv|I_g, 5f_{5/2}\rangle$ (with decay rates $\gamma_{56}$) are considered. Hyperfine structures of $|5\rangle$ and $|6\rangle$ are ignored, as they do not affect the isomeric population transfer from $\ket{1}$ to $\ket{3}$. $\Delta = E_2 - E_1 - \omega_e$ and $\delta = E_3 - E_1 - \omega_e - \omega_b$ denote the one-photon and two-photon detunings, respectively, where $\omega_e$ and $\omega_b$ are the frequencies of the electron-assisting and bridge-assisting  lasers.}}
\end{figure}

In $^{229}\mathrm{Th}^{3+}$ ions, the most efficient EB excitation involves the electronic transition $7p_{1/2}\rightarrow 7s_{1/2}$ \cite{PhysRevLett.105.182501}. Therefore, in this work, we investigate the hyperfine EB transition $[I_g, (7p_{1/2}) J=1/2, F_g]\rightarrow [I_e, (7s_{1/2} )J=1/2, F_e]$ driven by a bridge-assisting laser field. However, the $7p_{1/2}$ level is short-lived, with a lifetime of only about 1.1~ns \cite{Safronova2006}. In a practical scheme, both the preparation and decay of the $7p_{1/2}$ level must be considered. The $7s_{1/2}$ level is metastable with a lifetime of about 0.6~s \cite{Safronova2006}, so we assume the system is initially prepared in this level and an electron-assisting laser field is applied to drive the transition from $7s_{1/2}$ to $7p_{1/2}$. To account for the influence of the decay of $7p_{1/2}$, the $6d_{3/2}$ and $5f_{5/2}$ levels are also included. Consequently, a six-level model is constructed, as shown in Fig.~\ref{fig:2025_10_11_1}.

\subsection{Quantum Master Equation for Six-level Model in the $^{229}\mathrm{Th}^{3+}$ Ion}
The Hamiltonian of the six-level model in the interaction picture is written as [the atomic unit ($\hbar=|e|=m_e=$1) is used]
\begin{equation}
H(t) =\Delta \hat{ \sigma}_{22}
+\delta\hat{ \sigma}_{33}
+\left[\Omega_{e}(t)\hat{ \sigma}_{21}
+\Omega_{b} (t)\hat{ \sigma}_{32}+\text{H.c.}\right],
\label{eq:h}
\end{equation}
where $\Omega_{e}(t)$ and $\Omega_{b}(t)$ represent Rabi frequencies of the electron-assisting and bridge-assisting laser pulses, respectively. $\sigma_{ji}$ denotes the projector $|j\rangle\langle i|$ ($i,j=1,\cdots,6$). The one-photon and two-photon detunings are defined by $\Delta = E_2 - E_1 - \omega_e$ and $\delta = E_3 - E_1 - \omega_e-\omega_b$, where $\omega_e$ and $\omega_b$ are the frequencies of the electron-assisting and bridge-assisting lasers. Here, $E_j$ is the energy of $\ket{j}$. 
The time evolution of the system is described by the quantum master equation
\begin{equation}
\dot{\rho}= -i [H(t), \rho] + \mathcal{L}\rho,
\label{eq:masterE}
\end{equation}
where $\mathcal{L}\rho$ represents the Lindblad term given by
\begin{equation}
\mathcal{L}\rho=\frac{\gamma_{21}}{2} \mathcal{L}_{\hat{\sigma}_{12}}\rho+\frac{\gamma_{24}}{2} \mathcal{L}_{\hat{\sigma}_{42}}\rho+\frac{\gamma_{25}}{2} \mathcal{L}_{\hat{\sigma}_{52}}\rho+\frac{\gamma_{56}}{2} \mathcal{L}_{\hat{\sigma}_{65}}\rho.
\end{equation}
Here, $\gamma_{ij}$ ($i,j=1,\cdots,6$) denotes the decay rate from state $|i\rangle$ to state $|j\rangle$, and the operator $\mathcal{L}_{\hat{\sigma}_{ji}}\rho$ is defined as
\begin{equation}
\mathcal{L}_{\hat{\sigma}_{ji}}\rho=2\hat{\sigma}_{ji}\rho\hat{\sigma}_{ji}^{\dagger}-
\hat{\sigma}_{ji}^{\dagger}\hat{\sigma}_{ji}\rho-\rho\hat{\sigma}_{ji}^{\dagger}\hat{\sigma}_{ji}.
\label{eq:L}
\end{equation}
For the atomic electric-dipole transition between hyperfine levels $(I_g,\Gamma_iJ_i,F_i )$ and $(I_g,\Gamma_fJ_f,F_f )$, the decay rate $\gamma_{ij}$ is expressed as
\begin{equation}
\gamma_{ij} = \frac{4}{3}\left(\frac{\omega_{ij}}{c}\right)^{3}(2F_f+1) 
\left\{\begin{matrix}
J_f & J_i & 1 \\
F_i & F_f & I_g
\end{matrix}\right\}^2  \left|\left\langle\Gamma_{f}J_{f}\left|\left|\mathcal{O}^{(E1, \mathrm{length})}\right|\right|\Gamma_{i}J_{i}\right\rangle\right|^2,
\label{eq:gamma}
\end{equation}
where $\omega_{ij}$ is the transition energy, and $\mathcal{O}^{(E1, \mathrm{length})}=-rC^{(1)}(\theta,\phi)$ is the electric-dipole momentum operator in the length gauge. If the hyperfine structure is neglected, the decay rate between electronic levels $(\Gamma_iJ_i)$ and $(\Gamma_fJ_f)$ is obtained by summing  Eq.~\eqref{eq:gamma} over $F_f$:
\begin{equation}
\gamma_{ij}=  \frac{4}{3}\left(\frac{\omega_{ij}}{c}\right)^{3}
\frac{\left|\left\langle \Gamma_{f}J_{f}\left|\left|\mathcal{O}^{(E1, \mathrm{length})}\right|\right|\Gamma_{i}J_{i}\right\rangle\right|^2}{2J_i+1}.
\label{eq:gamma2}
\end{equation}
The atomic decay rates used in this work are $\gamma_{21}/2\pi = 35\, \text{MHz}$, $\gamma_{24}/2\pi = 28\, \text{MHz}$, $\gamma_{25}/2\pi = 100\, \text{MHz}$, and $\gamma_{56}/2\pi = 0.18\, \text{MHz}$~\cite{Wang2024-2}.

The Rabi frequencies $\Omega_b(t)$ and $\Omega_e(t)$ are given by
\begin{equation}
	\Omega_b(t)=\frac{E_{0,b}}{2}f_b(t)(-1)^{m }
\left(\begin{matrix}
	2  & 1   &  3 \\
	-m   & 0   & m
\end{matrix}\right)M_b,
\label{eq:OmegaB4}
\end{equation}
and
 \begin{equation}
 		\Omega_e(t) =- \frac{E_{0,e}}{2} f_e(t) (-1)^{  m} 
 	\left( \begin{matrix}
 		2& 1 & 3\\
 		-m & 0 & m
 	\end{matrix} \right) M_e,
 	\label{eq:OmegaE}
\end{equation}
respectively.  Here, $m$ denotes the magnetic quantum number, $E_{0,b}$ and $E_{0,e}$ are the strengths of the bridge-assisting and electron-assisting laser fields.
$f_b(t)$ and $f_e(t)$ represent the envelop functions of the bridge-assisting and electron-assisting laser fields. Both laser fields are assumed to be linearly polarized along the z axis.

The notation $M_b$ in Eq.~\eqref{eq:OmegaB4} denotes the reduced matrix element associated with the transition $\ket{2}\rightarrow\ket{3}$. The general expression of $M_b$ for the transition from the hyperfine level $(I_g, \Gamma_i J_i, F_i)$ to the hyperfine level $(I_e, \Gamma_fJ_f, F_f)$ is 
\begin{equation} 
	M_b=  \sqrt{[F_i][F_f]} (-1)^{I_g+F_i+J_f} \sum_{\tau K}\big[{M^{(\tau K,E1)}+M^{(E1,\tau K)}}\big]\left\langle I_e \left|\left|\mathcal{M}^{(\tau K)}\right|\right|I_g\right\rangle,
	\label{eq:OmegaB3}
\end{equation}
with the notation $[F]\equiv2F+1$ and the components:
\begin{equation}
	\begin{split}
		M^{(\tau K, E1)} = &
		\sum_t (-1)^{2I_g + 1 + J_i + J_f + 2F_f}
		\left\{ \begin{matrix}
			J_i & J_t & 1 \\
			F_f & F_i & I_g
		\end{matrix} \right\}
		\left\{ \begin{matrix}
			I_g & J_t & F_f \\
			J_f & I_e & K
		\end{matrix} \right\} \\
		& \times \left\langle \Gamma_f J_f \left| \left| T^{(\tau K)} \right| \right| \Gamma_t J_t \right\rangle
		\frac{\left\langle \Gamma_t J_t \left| \left| \mathcal{O}^{(E1, \mathrm{length})} \right| \right| \Gamma_i J_i \right\rangle}
		{E_{(I_e, \Gamma_f J_f)} - E_{(I_g, \Gamma_t J_t)}}, \\
		M^{(E1, \tau K)} = &
		\sum_t (-1)^{I_e + I_g + 1 + 2J_t + F_i + F_f}
		\left\{ \begin{matrix}
			J_t & J_f & 1 \\
			F_f & F_i & I_e
		\end{matrix} \right\}
		\left\{ \begin{matrix}
			I_e & J_t & F_i \\
			J_i & I_g & K
		\end{matrix} \right\} \\
		& \times \left\langle \Gamma_f J_f \left| \left| \mathcal{O}^{(E1, \mathrm{length})} \right| \right| \Gamma_t J_t \right\rangle
		\frac{\left\langle \Gamma_t J_t \left| \left| T^{(\tau K)} \right| \right| \Gamma_i J_i \right\rangle}
		{E_{(I_g, \Gamma_i J_i)} - E_{(I_e, \Gamma_t J_t)}}.
	\end{split}
	\label{eq:M}
\end{equation}
The detailed derivation of Eq.~\eqref{eq:OmegaB3} can be found in the Supplemental Material of Ref.~\cite{Wang2024-2}. By substituting the configurations of $\ket{2}$ and $\ket{3}$ into Eq.~\eqref{eq:OmegaB3}, the corresponding form of $M_b$ in Eq.~\eqref{eq:OmegaB4} is obtained.

Similarly, the reduced matrix element \(M_e\) in Eq.~\eqref{eq:OmegaE} is associated with the transition $\ket{1}\rightarrow\ket{2}$, 
and its general expression for the transition $(I_g, \Gamma_i J_i, F_i)\rightarrow(I_e, \Gamma_fJ_f, F_f)$ is given by
\begin{equation}
	M_e = (-1)^{I_g + 1 + J_i + F_f} \sqrt{[F_i][F_f]} 
	\left\{ \begin{matrix}
		J_f & J_i & 1 \\
		F_i & F_f & I_g
	\end{matrix} \right\} 
	\left\langle\Gamma_f J_f\left|\left|\mathcal{O}^{(E1, \mathrm{length})}\right|\right|\Gamma_i J_i\right\rangle.
	\label{eq:OmegaE2}
\end{equation}
Note that we have corrected the sign terms in Eqs.~\eqref{eq:M} and \eqref{eq:OmegaE2}, which were given incorrectly in Ref.~\cite{Wang2024-2}. Fortunately, the influence of these errors on the numerical results in Ref.~\cite{Wang2024-2} is negligible.

\subsection{Shortcut Design}
In the case of two-photon resonance $(\delta = 0)$, Ref.~\cite{Wang2024-2} have achieved the isomeric population transfer by using the STIRAP scheme for the six-level system of  $^{229}\mathrm{Th}^{3+}$ ions. In that scheme, it involves the time-dependent but slowly varying Rabi frequencies $\Omega_b(t)$ and $\Omega_e(t)$, allowing the system to remain in the dark state 
\begin{equation}
    |D(t)\rangle=\frac{1}{\sqrt{\Omega_b^2(t)+\Omega_e^2(t)}}\left[\Omega_b(t)|1\rangle-\Omega_e(t)|3\rangle\right]
\label{eq:dark state}
\end{equation}
throughout the entire adiabatic process, thereby realizing population transfer from the initial state $|1\rangle$ to the target state $|3\rangle$. This method maintains stable performance despite small variations in laser parameters, such as intensity and detuning.

To realize swift population transfer from \( |1\rangle \) to \( |3\rangle \) via STIRAP within a reduced timescale, one can employ counterdiabatic driving under the two-photon resonance condition~\cite{PhysRevLett.105.123003,Torrents2013,Berry2009,Demirplak2003,Demirplak2005}, which is consistent with the quantum transitionless algorithm. This technique introduces an auxiliary interaction expressed as
\begin{equation}
H_{\text{cd}}(t) = i\hbar \sum_{n} \left( |\partial_t n\rangle \langle n| - \langle n|\partial_t n\rangle |n\rangle \langle n| \right),
\label{eq:Hcd}
\end{equation}
where \( |n\rangle \) denotes the instantaneous eigenstates of \( H(t) \). The Hamiltonian \( H_{\text{cd}}(t) \) can be tailored specifically as
\begin{equation}
H_{\text{cd}} = i\hbar \Omega_a(t) |1\rangle\langle 3| + \text{H.c.},
\label{eq:Hcd2}
\end{equation}
with the coefficient \( \Omega_a(t) \) defined by \( \Omega_a(t) = [\dot{\Omega}_e(t) \Omega_b(t) - \dot{\Omega}_b(t) \Omega_e(t)] / [\Omega_e^2(t) + \Omega_b^2(t)] \). This additional coupling between \( |1\rangle \) and \( |3\rangle \), enabled by a dipole transition in an atomic setup, effectively mitigates non-adiabatic transitions.

When the detuning of the intermediate level becomes substantial (\(\Delta \gg \Omega_e^{\max}, \Omega_b^{\max}\)), the population in state \(|2\rangle\) remains negligible (\(\dot{c}_2(t) \approx 0\)), allowing for its adiabatic elimination by neglecting fast-oscillating off-diagonal terms. This procedure yields an effective two-level Hamiltonian within the \(\{|1\rangle, |3\rangle\}\) subspace~\cite{PhysRevA.55.648}:
\begin{equation}
H_{\text{eff}} = \frac{1}{2} 
\begin{pmatrix} 
-\frac{|\Omega_e(t)|^2 - |\Omega_b(t)|^2}{\Delta} & -\frac{2\Omega_e(t) \Omega_b(t)}{\Delta} \\ 
-\frac{2\Omega_e(t) \Omega_b(t)}{\Delta} & \frac{|\Omega_e(t)|^2 - |\Omega_b(t)|^2}{\Delta}
\end{pmatrix},
\label{eq:Heff}
\end{equation}
With this effective two-level description, the counterdiabatic driving term can be systematically derived based on the time evolution of the effective Hamiltonian~\cite{PhysRevLett.105.123003}:
\begin{equation}
H_{\text{cd}} = \frac{1}{2} 
\begin{pmatrix} 
0 & i\alpha(t) \\ 
-i\alpha(t) & 0 
\end{pmatrix},
\label{eq:Hcd3}
\end{equation}
where \(\alpha(t)\) is derived from the time derivatives of the effective Rabi frequency and detuning, 
\(\alpha(t) = [\dot{\Omega}_{\text{eff}}(t) \Delta_{\text{eff}}(t) - \Omega_{\text{eff}}(t) \dot{\Delta}_{\text{eff}}(t)]/[\Delta_{\text{eff}}^2(t) + \Omega_{\text{eff}}^2(t)]\) with \(\Delta_{\text{eff}}(t) = (|\Omega_e(t)|^2 - |\Omega_b(t)|^2)/\Delta\) and \(\Omega_{\text{eff}}(t) = -2\Omega_e(t) \Omega_b(t)/\Delta\). The complete driving Hamiltonian becomes \(H_{\text{total}} = H_{\text{eff}} + H_{\text{cd}}\). By applying certain unitary transformations~\cite{PhysRevA.94.063411,PhysRevA.93.052324,PhysRevLett.116.230503}, \(H_{\text{total}}\) can be recast into 
\begin{equation}
\tilde{H}_{\text{total}} = \frac{1}{2} 
\begin{pmatrix} 
-\frac{|\tilde{\Omega}_e(t)|^2 - |\tilde{\Omega}_b(t)|^2}{\Delta} & -\frac{2\tilde{\Omega}_e(t) \tilde{\Omega}_b(t)}{\Delta} \\ 
-\frac{2\tilde{\Omega}_e(t) \tilde{\Omega}_b(t)}{\Delta} & \frac{|\tilde{\Omega}_e(t)|^2 - |\tilde{\Omega}_b(t)|^2}{\Delta}
\end{pmatrix}
\label{eq:Htal}
\end{equation}
with
\begin{equation}
\begin{aligned}
\tilde{\Omega}_e(t) &= \left\{\frac{\Delta}{2}
\left[\left(\tilde{\Delta}_{\text{eff}}^2(t) + \tilde{\Omega}_{\text{eff}}^2(t)\right)^{\frac{1}{2}} + \tilde{\Delta}_{\text{eff}}(t) \right]\right\}^{\frac{1}{2}}, \\
\tilde{\Omega}_b(t) &= \left\{\frac{\Delta}{2} \left[\left(\tilde{\Delta}_{\text{eff}}^2(t) + \tilde{\Omega}_{\text{eff}}^2(t)\right)^{\frac{1}{2}} - \tilde{\Delta}_{\text{eff}}(t) \right]\right\}^{\frac{1}{2}}, 
\end{aligned}
\label{eq:modify}
\end{equation}
where the positive and negative signs in the square roots are chosen to ensure \(\tilde{\Omega}_e(t)\) and \(\tilde{\Omega}_b(t)\) correspond to the electron-assisting and bridge-assisting field amplitudes, respectively. We note that $\Delta$ appearing in Eqs.~(\ref{eq:Htal},\ref{eq:modify}) should have an additional modulation. However, such a modulation can be negligible~\cite{PhysRevA.94.063411} and has been neglected. This suggests that, upon reverting to the full three-level description, we can engineer modified electron-assisting and bridge-assisting laser pulses to replicate the transformed effective Hamiltonian, ensuring consistency with the two-level approximation.

\begin{figure}[t!]
\centering
\includegraphics[width=13cm] {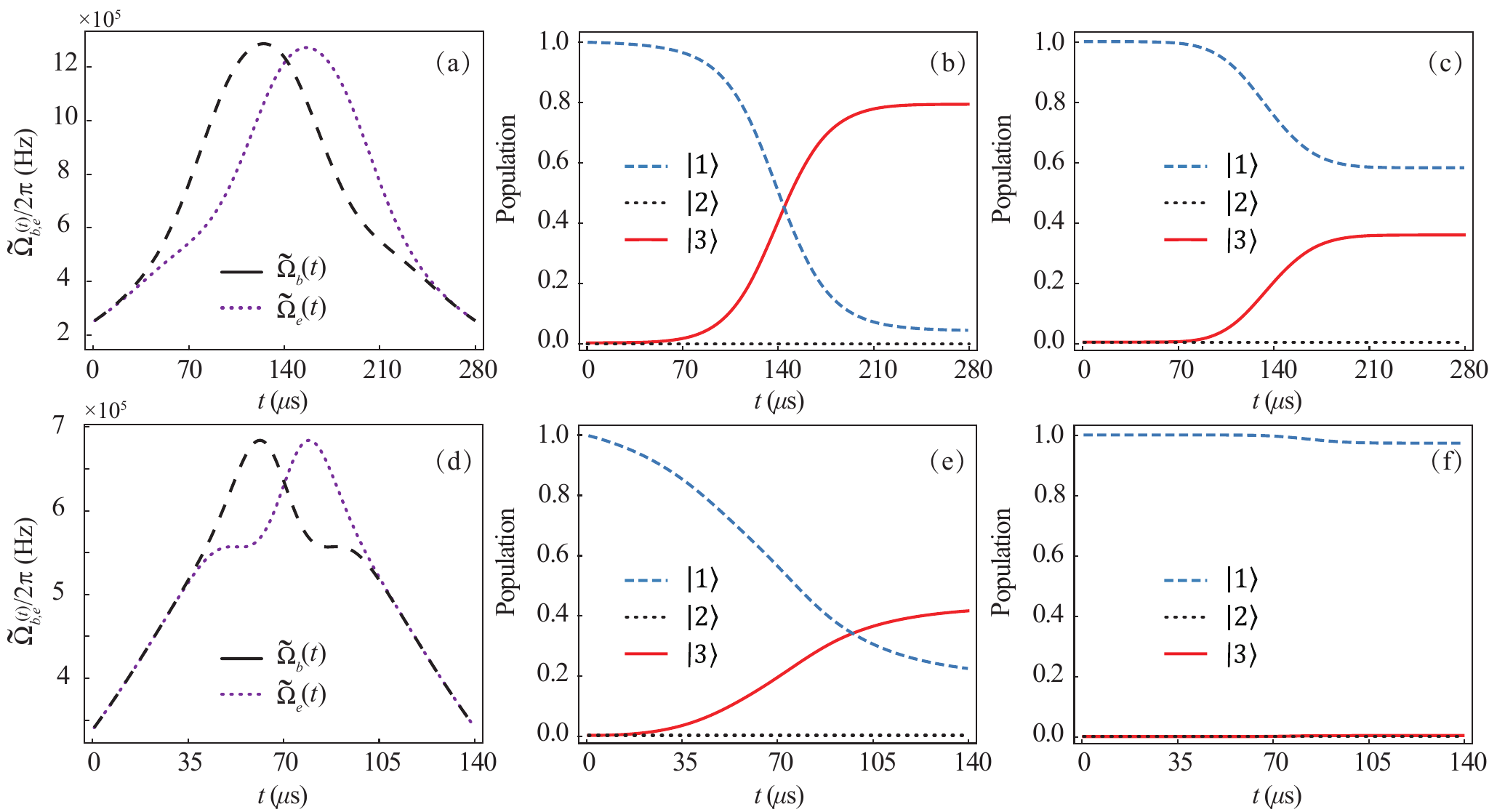}\\[5pt]  
\parbox[c]{15cm}{\footnotesize{\bf Fig.~2.} The upper and lower panels correspond to total operation times of \(t_{\text{total}} = 280 \, \mu\text{s}\) and \(t_{\text{total}} = 140 \, \mu\text{s}\), respectively. (a,\,d) Rabi frequencies \(\tilde{\Omega}_e(t) \) and \( \tilde{\Omega}_b(t)\) employed in the STIRSAP scheme. (b,\,e) Population transfer efficiencies from state \(|1\rangle\) to state \(|3\rangle\) achieved via the STIRSAP scheme, which are \( 79.38\% \) and \( 35.63\% \), respectively. (c,\,f) Population transfer efficiencies obtained with the STIRAP scheme, corresponding to \(41.88\%\) and \(0.36\%\), respectively.
The parameters are: (a,\,b) $\sigma = 54\,\mu\text{s}$, $t_e = 120\,\mu\text{s}$, $t_b = 160\,\mu\text{s}$, $\Omega_b/2\pi = \Omega_e/2\pi = 6.8 \times 10^5\,\text{Hz}$, $\Delta = 1000 |\Omega_b(t_b)|$; (c) $\sigma = 67\,\mu\text{s}$, $t_e = 120\,\mu\text{s}$, $t_b = 122\,\mu\text{s}$, $\Omega_b/2\pi = \Omega_e/2\pi = 12.9 \times 10^5\,\text{Hz}$, $\Delta = 1000 |\Omega_b(t_b)|$; (d,\,b) $\sigma = 20\,\mu\text{s}$, $t_e = 67\,\mu\text{s}$, $t_b = 73\,\mu\text{s}$, $\Omega_b/2\pi = \Omega_e/2\pi = 1.8 \times 10^5\,\text{Hz}$, $\Delta = 1000 |\Omega_b(t_b)|$; (f) $\sigma = 23\,\mu\text{s}$, $t_e = 56\,\mu\text{s}$, $t_b = 84\,\mu\text{s}$, $\Omega_b/2\pi = \Omega_e/2\pi = 1.8 \times 10^5\,\text{Hz}$, $\Delta = 1000 |\Omega_b(t_b)|$. 
Throughout all subplots, two-photon detuning is zero $(\delta = 0)$, and the atomic decay rates are $\gamma_{21}/2\pi = 35\, \text{MHz}$, $\gamma_{24}/2\pi = 28\, \text{MHz}$, $\gamma_{25}/2\pi = 100\, \text{MHz}$, and $\gamma_{56}/2\pi = 0.18\, \text{MHz}$.}
\end{figure}

\section{Realization of STIRSAP}
Within the STIRAP method, the envelope functions of the bridge-assisting and electron-assisting laser fields are modeled as slowly varying Gaussian pulses:
\begin{equation}
\begin{aligned}
f_{e}(t) &= \exp\left[-\frac{(t - t_e)^2}{\sigma^2}\right], \\
f_{b}(t) &= \exp\left[-\frac{(t - t_b)^2}{\sigma^2}\right].
\end{aligned}
\label{eq:original}
\end{equation}

Here, $\sigma$ represents the full width at half-maximum, while $t_e$ and $t_b$ denote the center times of the respective pulses. To satisfy the adiabatic condition, which is crucial for high-fidelity population transfer, these pulses must evolve slowly, necessitating sufficiently long durations and a significant temporal overlap. As a benchmark, we consider the STIRAP implementation in Ref.~\cite{Wang2024-2}. In that work, the laser intensity $I_0$ ($\equiv\epsilon_0 E_{0,b}^{2}/2$ with vacuum permittivity $\epsilon_0$) of the bridge-assisting laser field is set to $2\times 10^7\,\text{ W/cm}^2$, corresponding to the peak Rabi frequency $\Omega_b(t_b)/2\pi = 6.8\times 10^5 \,\text{Hz}$. The pulse parameters were set to $\sigma=18\,\text{ms}$, $t_b=55\,\text{ms}$, and $t_e=82\,\text{ms}$, with a total operation time of $t_{\text{total}} = 140\,\text{ms}$. This configuration achieved a remarkable final transfer efficiency of approximately $99.7\%$.

In STIRSAP scheme, based on Eq.~(\ref{eq:modify}), the modified electron-assisted and bridge-assisted laser pulses replace the original pulses in Eq.~(\ref{eq:original}). The peak Rabi frequencies, before any modification,  are set as \(\Omega_b(t_b)/2\pi = \Omega_e(t_e)/2\pi=6.8 \times 10^5\,\text{Hz}\). Let us first consider the case of two-photon resonance ($\delta = 0$), and the one-photon detuning is chosen as \(\Delta = 1000\, |\Omega_b(t_b)|\). By optimizing the pulse parameters to \(\sigma = 54 \, \mu\text{s}\), \(t_b = 120\, \mu\text{s}\), and \(t_e = 160 \, \mu\text{s}\), the resulting Rabi frequencies of the modified pulses, $\tilde{\Omega}_b(t)/2\pi$ and $\tilde{\Omega}_e(t)/2\pi$, are shown in Fig.~2(a), whose peak values are $12.9\times 10^5\,\text{Hz}$. 
The total operation time is \(t_{\text{total}} = 280\, \mu\text{s}\), which is merely 1/500 of the time required for STIRAP in Ref.~\cite{Wang2024-2}. Under these conditions, the population transfer efficiency reaches 79.38\% [see Fig.~2(b)].

The modified pulses exhibit higher peak intensities and different pulse durations compared with the original pulses. To ensure a fair comparison, the peak Rabi frequencies and durations of the pulses in the STIRAP scheme were matched to those in the STIRSAP scheme. Therefore, the parameters of the STIRAP scheme were optimized accordingly as $\sigma = 67\, \mu\text{s}$, $t_b = 120\, \mu\text{s}$, and $t_e = 122 \, \mu\text{s}$, with the same peak Rabi frequency $\Omega_b(t_b)/2\pi =12.9\times 10^5 \,\text{Hz}$ and a total operation time of $t_{\text{total}} = 280\, \mu\text{s}$. In this case, the STIRAP scheme achieved a population transfer efficiency of only $35.63\%$ from state $|1\rangle$ to state $|3\rangle$ [see Fig.~2\,(c)]. Compared to the STIRAP scheme, the STIRSAP scheme achieves higher population transfer efficiency while operating in a significantly reduced time.

To further explore the performance limits, we reduce the total operation time to \(t_{\text{total}} = 140\, \mu\text{s}\), a further reduction from the previous case. For the STIRSAP scheme, 
the peak Rabi frequencies, before any modification,  are set as \(\Omega_b(t_b)/2\pi = \Omega_e(t_e)/2\pi=1.8 \times 10^5\,\text{Hz}\). The detuning is \(\Delta = 1000\, |\Omega_b(t_b)|\), with the system under two-photon resonance ($\delta = 0$).
The Rabi frequencies of the modified pulses derived from Eq.~(\ref{eq:modify}), $\tilde{\Omega}_b(t)/2\pi$ and $\tilde{\Omega}_e(t)/2\pi$, are shown in Fig.~2(d), whose peak values are $6.8\times 10^5\,\text{Hz}$. The corresponding optimized pulse parameters are
\(\sigma = 20\, \mu\text{s}\), \(t_b = 67\, \mu\text{s}\), and \(t_e = 73\, \mu\text{s}\), achieving a population transfer efficiency of \(41.88\%\) [see Fig.~2(e)]. In contrast, the peak Rabi frequencies and half-widths of the pulses in the STIRAP scheme were matched to those in the STIRSAP scheme.
The optimized parameters are set as \(\sigma = 23\, \mu\text{s}\), \(t_b = 56\,\mu\text{s}\), and \(t_e = 84\, \mu\text{s}\), yet the population transfer efficiency is only \(0.36\%\) [see Fig.~2(f)]. This demonstrates the superior performance of the STIRSAP scheme under such compressed timescales.

\begin{figure}[t!]
\centering
\includegraphics[width=11cm] {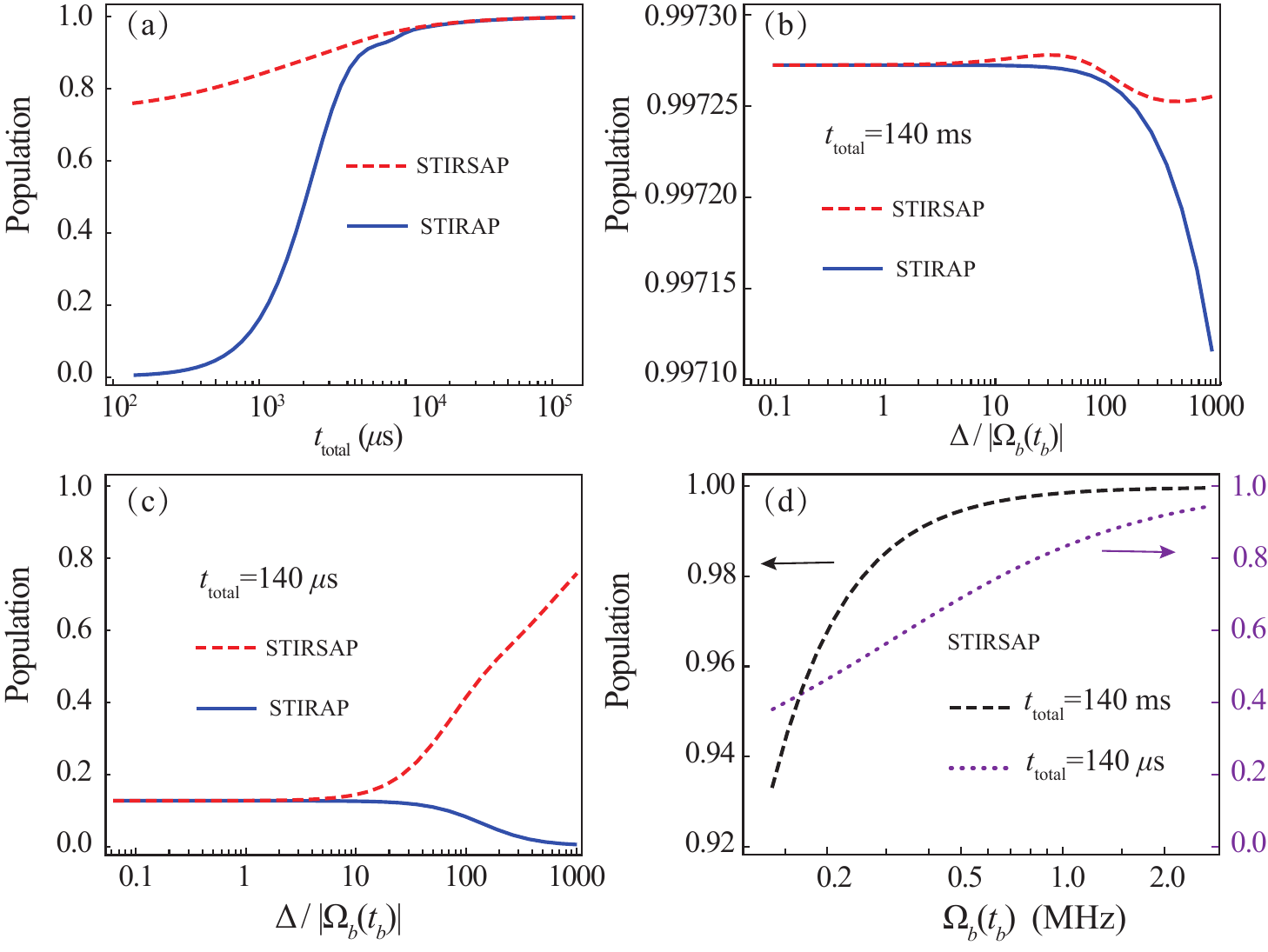}\\[5pt] 
\parbox[c]{15cm}{\footnotesize{\bf Fig.~3.} (a) Influence of the total operation time \(t_{\text{total}}\) on the population transfer efficiency of the STIRSAP scheme (red dashed line) and the STIRAP scheme (blue solid line), respectively. Parameters are set as \(\Omega_b(t_b)/2\pi = \Omega_e(t_e)/2\pi=6.8 \times 10^5\,\text{Hz}\) and \(\Delta = 1000\, |\Omega_b(t_b)|\). (b,\,c) Influence of the detuning $\Delta$ on the population transfer efficiency, comparing the STIRSAP (red dashed line) and the STIRAP scheme (blue solid line).  Parameters are set as \(\Omega_b(t_b)/2\pi = \Omega_e(t_e)/2\pi=6.8 \times 10^5\,\text{Hz}\). The total operation times are $140\,\text{ms}$ for (b) and $140\,\mu\text{s}$ for (c), respectively. (d) Influence of the peak Rabi frequencies $\Omega_b(t_b)=\Omega_e(t_e)$ on the population transfer efficiency of the STIRSAP scheme for long operation times $t_{\text{total}} = 140\,\text{ms}$ (black dashed line) and short operation times $t_{\text{total}} = 140\,\mu\text{s}$ (purple dotted line). The detuning is \(\Delta = 1000\, |\Omega_b(t_b)|\). All subplots are under fixed Gaussian pulse shapes with $\sigma = t_{\text{total}}/6$, $t_e = 2\,t_{\text{total}}/5$, and $t_b = 3\,t_{\text{total}}/5$. The two-photon detuning is zero $(\delta = 0)$, and the atomic decay rates are the same as those in Fig.~2.}
\end{figure}

To elucidate the advantages of the STIRSAP scheme over the STIRAP scheme for population transfer in the $^{229}\mathrm{Th}^{3+}$ ions, we systematically compare their performance in the total operation time, the detuning, and the peak Rabi frequency  under the condition of two-photon resonance. These comparisons highlight the advantages of STIRSAP, which benefits from the counter-diabatic driving.
First, under identical conditions of the peak Rabi frequencies \(\Omega_b(t_b) = \Omega_e(t_e)\), the detuning \(\Delta\), and fixed Gaussian pulse shapes, the total operation time \(t_{\text{total}}\) significantly affects the population transfer efficiency. As shown in Fig.~3(a), reducing \(t_{\text{total}}\) from $140\,\text{ms}$ to $140\,\mu\text{s}$ leads to a decrease in population transfer efficiency from nearly $100\%$ to $75.91\%$ in the STIRSAP scheme (red dashed line). In contrast, the population transfer efficiency of the STIRAP scheme drops sharply to only $6.78\%$ (blue solid line).

Second, under identical conditions of the peak Rabi frequencies \(\Omega_b(t_b) = \Omega_e(t_e)\), the total operation time $t_{\text{total}}$, and fixed Gaussian pulse shapes, the detuning \(\Delta\) affects the two schemes differently and varies with the operation time. 
For a long time (\(t_{\text{total}} = 140\,\text{ms}\)), as shown in Fig.~3(b), both schemes maintain efficiencies above $99.7\%$ in various ranges \(\Delta\). However, under large detuning, the population transfer efficiency of STIRSAP (red dashed line) slightly outperforms that of STIRAP (blue solid line). 
For a short time (\(t_{\text{total}} = 140\,\mu\text{s}\)), as depicted in Fig.~3(c), the behaviors are markedly different. For small detuning, efficiencies are comparable; for large detuning, the population transfer efficiency of the STIRSAP scheme (red dashed line) increases with detuning, while the one of the STIRAP scheme (blue solid line) decreases slightly. 
This enhancement of the STIRSAP scheme arises from the role of detuning in counteracting dissipation effects, effectively protecting population transfer from intermediate-state decay (the state $|2\rangle$).

Under identical conditions of the detuning \(\Delta\), the total operation time $t_{\text{total}}$, and fixed Gaussian pulse shapes,  the peak Rabi frequencies \(\Omega_b(t_b) = \Omega_e(t_e)\) affect the population transfer on the STIRSAP scheme. 
For a long time (\(t_{\text{total}} = 140\,\text{ms}\)), the increasing of $\Omega_b(t_b)$ results in a slight increase efficiency (from \(93.45\%\) to \(99.97\%\)) [see the black dotted line in Fig.~3(d)].
For a short time (\(t_{\text{total}} = 140\,\mu\text{s}\)), as depicted by the purple dashed line in Fig.~3(d), the efficiency rises significantly from \(38.46\%\) to \(94.13\%\) with the increasing of $\Omega_b(t_b)$. Thus, we must fully consider the influence of enhanced pulse intensity, which is also the reason for conducting comparisons at the same pulse intensity in Fig.~2.

Finally, in Fig.~4, we further examine the influence of two-photon detuning $\delta$ on the isomeric population transfer, fixing the one-photon detuning at $\Delta = 1000 \Omega_b$. For comparison, the corresponding result for STIRAP is also shown. 
For the STIRSAP scheme (red dashed line), when the two-photon detuning reaches about $3.9\times 10^{-3}\, \Omega_b$ (i.e., $\delta/2\pi\approx2.7\times10^3$ Hz), the isomeric population transfer decreases to half of its maximum value, reaching about 39.69\%. Beyond this threshold, the isomeric population transfer drops rapidly to zero. In contrast, the STIRAP scheme (blue solid line) is more sensitive to $\delta$: the corresponding half-maximum threshold occurs at about  $2.7\times 10^{-3}\, \Omega_b$, which is smaller by a factor of $1.4$ compared with the STIRSAP scheme. Consequently, STIRSAP is more experimentally advantageous in the presence of measurement uncertainties in nuclear or electronic energy levels.

These results collectively demonstrate that STIRSAP not only accelerates population transfer by orders of magnitude but also maintains high efficiency over a wide range of detuning, making it particularly suitable for nuclear quantum optics applications where dissipation and time constraints are critical.

\begin{figure}[t!]
\centering
\includegraphics[width=5cm] {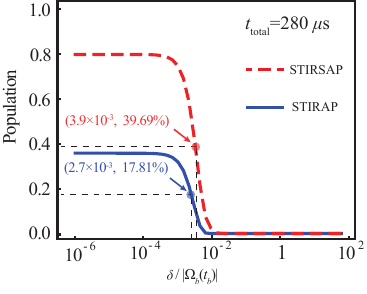}\\[5pt]  
\parbox[c]{15cm}{\footnotesize{\bf Fig.~4.} Isomeric population tranfer efficiency from state $|1\rangle$ to state $|3\rangle$ as a function of the two-photon detuning $\delta$ for the STIRSAP (red dashed line) and the STIRAP (blue solid line) scheme. The total operation time is $280\,\mu\text{s}$. The Gaussian pulse shapes are  $\sigma = 54\,\mu\text{s}$, $t_e = 120\,\mu\text{s}$, $t_b = 160\,\mu\text{s}$, $\Omega_b/2\pi = \Omega_e/2\pi = 6.8 \times 10^5\,\text{Hz}$ for STIRSAP scheme, and  $\sigma = 67\,\mu\text{s}$, $t_e = 120\,\mu\text{s}$, $t_b = 122\,\mu\text{s}$, $\Omega_b/2\pi = \Omega_e/2\pi = 12.9 \times 10^5\,\text{Hz}$ for STIRAP scheme. The detuning is \(\Delta = 1000\, |\Omega_b(t_b)|\). The atomic decay rates are the same as those in Fig.~2.}
\end{figure}

\section{Conclusion}
In summary, we have demonstrated that the STIRSAP method provides a rapid and efficient approach for achieving isomeric population transfer in the $^{229}$Th$^{3+}$ ions. Compared with the STIRAP method, STIRSAP significantly shortens the operation time from the order of hundreds of milliseconds to hundreds of microseconds, while maintaining a high transfer efficiency of $79.38\%$. Even under shorter pulse durations, where the STIRAP method becomes almost ineffective, the STIRSAP method still achieves an isomeric population transfer exceeding 40\%. Compared with STIRAP, the STIRSAP method is  more robust against uncertainties in the two-photon detuning, which is particularly advantageous in the presence of uncertainties in nuclear or electronic energy levels.

\section*{Acknowledgments}
This work was supported by the National Natural Science Foundation of China (Grants No.~12574387, No.~12405026 and No.~12274107), the Innovation Program for Quantum Science and Technology (Grant No.~2023ZD0300704), and the Funds of the Natural Science Foundation of Hangzhou (Grant No.~2024SZRY\\BA050001) and the Strategic Priority Research Program of the Chinese Academy of Sciences (Grant No. XDB0920000).

\bibliographystyle{unsrt}  
\bibliography{references}

\end{document}